\RequirePackage{fix-cm}
\documentclass{svjour3}                     
\smartqed  
\usepackage[export]{adjustbox}
\usepackage{graphicx,amsmath,amssymb,setspace,multirow}
\usepackage{setspace}
\numberwithin{equation}{section}
\usepackage{multirow}
\newcommand{\pard}[2]{\frac{\partial #1}{\partial #2}}
\usepackage[table]{xcolor}
\usepackage{booktabs}
\usepackage{url}
\usepackage{chemformula}
\definecolor{mygray}{gray}{0.6}
\begin{document}

\title{Numerical predictions of  shear stress and cyclic stretch in the healthy pulmonary vasculature
}
%

\allowdisplaybreaks
\titlerunning{Numerical predictions of  shear stress and cyclic stretch}        

\author{Michelle A. Bartolo  \and
        M. Umar Qureshi \and
        Mitchel J. Colebank \and 
        Naomi C. Chesler \and
        Mette S. Olufsen
}

\authorrunning{M.A. Bartolo} 

\institute{Michelle A. Bartolo \at
Department of Mathematics\\
           North Carolina State University\\
           Raleigh, NC 27695\\
           \email{mabartol@ncsu.edu}         
           \and
           M. Umar Qureshi \at
            Department of Mathematics\\
           North Carolina State University\\
           Raleigh, NC 27695\\
           \email{muquresh@ncsu.edu} 
           \and
Mitchel J. Colebank\at
           Department of Mathematics\\
           North Carolina State University\\
           Raleigh, NC 27695 \\
           \email{mjcoleba@ncsu.edu}  
           \and
          Naomi C. Chesler\at
           The Henry Samueli School of Engineering\\
University of California, Irvine\\
Irvine, CA 92697 \\
           \email{nchesler@uci.edu}  
           \and
           Mette S. Olufsen \at
           Department of Mathematics\\
           North Carolina State University\\
           Raleigh, NC 27607 \\
           \email{msolufse@ncsu.edu}  
}

\date{Received: date / Accepted: date}

\maketitle

\begin{abstract}
Isolated post-capillary pulmonary hypertension (Ipc-PH) occurs due to left heart failure, which contributes to 1 out of every 9 deaths in the United States. In some patients, through unknown mechanisms, Ipc-PH transitions to combined pre-/post-capillary PH (Cpc-PH), diagnosed by an increase in pulmonary vascular resistance and associated with a dramatic increase in mortality. We hypothesize that altered mechanical forces and subsequent vasoactive signaling in the pulmonary capillary bed drive the transition from Ipc-PH to Cpc-PH. However, even in a healthy pulmonary circulation, the mechanical forces in the smallest vessels (the arterioles, venules, and capillary bed) have not been quantitatively defined. This study is the first to examine this question via a computational fluid dynamics model of the human pulmonary arteries, veins, arterioles, and venules. Using this model we predict temporal and spatial dynamics of cyclic stretch and wall shear stress. In the large vessels, numerical simulations show that shear stress increases coincides with larger flow and pressure. In the microvasculature, we found that as vessel radius decreases, shear stress increases and flow decreases. In arterioles, this corresponds with lower pressures; however, the venules and smaller veins have higher pressure than larger veins.  Our model provides predictions for pressure, flow, shear stress, and cyclic stretch that provides a way to  analyze and investigate  hypotheses related to disease progression in the pulmonary circulation. 

\end{abstract}

\section{Introduction}\label{sec:intro}

\par Chronic left heart failure (LHF) impacts nearly 5.9 million adults and contributes to 1 out of 9 deaths in the United States \cite{mozaffarian2016LHF}. About 60-80\% of patients with LHF develop pulmonary hypertension (PH-LHF) \cite{ghio2001PH,lam2009PH}, which dramatically increases morbidity and mortality \cite{guazzi2012PHmorbidity,guglin2010PHmorbidity,ramu2016PHmorbidity}. According to the 2019 World Health Organization guidelines, diagnosis of PH-LHF (classified under group-II PH) \cite{simonneau2019PHdiagnosis} requires a mean pulmonary artery pressure (mPAP) $> 20$ mmHg and a pulmonary capillary wedge pressure (PCWP) $> 15$ mmHg\cite{simonneau2019PHdiagnosis}. The disease begins with isolated post-capillary PH (Ipc-PH), a passive process characterized by sustained elevated left-atrial and pulmonary venous pressures, leading to increases in pulmonary arterial pressure. The increased pressure progressively remodels and stiffens the vasculature, often starting in the small vessels and propagating to the main pulmonary artery (MPA) \cite{ravi2013PHprogeression}. In some patients, through unknown mechanisms, Ipc-PH transitions to combined pre-/post-capillary PH (Cpc-PH), which significantly worsens the disease prognosis. The increase in pulmonary vascular resistance (PVR) marks a change from reversible to the irreversible stage, significantly increasing the mortality risk \cite{gerges2015PHprogeression,miller2013PHprogression}. We hypothesize that altered mechanical forces and subsequent vasoactive signaling in pulmonary capillary beds drive the transition from Ipc-PH and Cpc-PH. However, mechanical forces in the small pulmonary arterioles and venules have not yet been quantitatively studied, even in the healthy pulmonary vasculature. In both healthy subjects and patients with PH-LHF, a crucial precursor to investigating mechanobiological mechanisms of disease progression is quantifying the dynamic vascular mechanical environment. Therefore, this study examines the time course and distribution  of wall shear stress (WSS) and cyclic stretch (CS) in the healthy pulmonary circulation, enhancing our understanding of how mechanical stimuli act on the non-diseased vasculature and providing a basis for studying the PH-LHF pathogenesis. 

\par WSS (g/cm$\cdot$s$^2$) is a potential quantifier of endothelial function and responsiveness for assessing PH \cite{moraes1997pulmonary}. It is defined as the tangential force exerted by the flowing blood per unit surface area of tunica intima (i.e., the interior surface blood vessels covered with a thin layer endothelial cells called the endothelium). WSS is a local quantity and not an integrated property of the entire intima, yet it is intimately connected to the momentum exchange at the surface. As such, WSS is directly proportional to blood flow and viscosity, and indirectly proportional to vessel radius cubed \cite{paszkowiak2003arterial}. Endothelial cell sensitivity to WSS is associated with vascular remodeling, which has a local impact on the pulmonary arteries and veins and may lead to the worsening of PH-LHF \cite{roux2020fluid}. It is also a predictor of endothelial function due to its role in stimulating the production, release, and penetration of nitric oxide (NO) from the intima to the smooth muscle cells in the media. NO acts as a vital protective agent that maintains hemodynamic homeostasis by regulating vascular wall mechanics \cite{Thomas2010}. Specifically, its production during periods of high flow rate leads to vasodilation, which maintains PVR within baseline values. However, as PH progresses, blood vessel permeability and distensibility decrease with the deteriorating endothelium. Endothelial dysfunction leads to a lower production of NO, creating an imbalance between WSS and vasodilation, and implicitly impacting the ability of noninvasively measured WSS to predict PH severity \cite{bleakley2015endothelial}.

\par CS (\%) is another physiological quantity that plays a role in endothelial function. Decreased CS in endothelial cells stimulates vasoconstriction via increased endothelin (ET-1) and decreased endothelial nitric oxide synthase (eNOS) \cite{van2019transition}. In vascular smooth muscle cells, CS causes cell cycle arrest in vascular smooth muscle cells by elongating and realigning cells in the direction of stretch \cite{barron2007effect}. Combined with increased blood pressure and transmural stress, these changes result in thickening of the vascular wall, increased wall thickness, increased media-to-lumen ratio, and decreased lumen diameter.  The combination of these multiple alternations to the vascular structure leads to increases in PVR and decreases in the compliance of arteries. \cite{birukov2009cyclic}.  Thus, understanding the role of WSS and CS within the pulmonary circulation and the relationship between these two vascular mechanical forces may give insight into the pathogenesis of PH and other cardiovascular diseases.

\par It is difficult to approximate WSS and CS in vivo as the measurable hemodynamic quantities cannot quantitatively define shearing forces or separate their effects from the other physical parameters of the hemodynamic system \cite{papadaki1999quantitative}. However, computational fluid dynamics (CFD) provides a comprehensive theoretical framework for determining WSS magnitude and time course. Previous CFD models (e.g., \cite{davies2009hemodynamic,reymond2011validation,reymond2012patient}) compute two WSS markers: the time averaged WSS (TAWSS) and an oscillatory shear stress index (OSI), not quantifying CS or differentiating the response between large and small vessels. These detailed predictions are excellent at examining the global environment, yet do not predict microcirculatory and venous WSS or CS and their relationships with large arterial hemodynamics. 

\par In this study, we develop a one-dimensional fluid dynamic model of the pulmonary circulation including both arteries and veins. We use a numerical algorithm to predict pressure drop, wave propagation,and magnitude and time course of  WSS, and CS  \cite{qureshi2014numerical}. The model includes large and small blood vessels capturing vessels with radii from --cm to --$\mu$m. This variation in spatial scale allows quantification and exploration of system dynamics and mechanistic relationships (e.g., between microcirculatory and large arterial WSS or CS) at several levels of the pulmonary circulation. \cite{selzer1992understanding}. 
 
\par Our model predicts these quantities in a network including large arteries and veins extracted from a computed tomography (CT) image from a healthy human subject. This allows us to study the distal vasculature in detail by predicting pressure, flow, WSS, and CS waveforms throughout the simulated network.  To compare our results with  3D studies, we also compute a 1D-TAWSS and 1D-OSI by averaging WSS and subsequently dividing by the amplitude of WSS. Investigation of these quantities in healthy pulmonary vasculature is an essential step for providing an understanding of disease mechanisms in PH-LHF patients.

\begin{table}[ht!]
\caption{List of parameters and variables.}
    \label{tab:defn}
    \footnotesize
    \centering
\rowcolors{2}{black!15}{white!50}
\begin{center}
\begin{tabular}{ c l l l }    \toprule
 \textbf{Quantity}  & \textbf{Definition} & \textbf{Units} & \textbf{Value}\\
\midrule
\textbf{Parameters} & & &\\
\midrule
  $T$ & Length of the cardiac cycle & s & $1$\\
$\rho$ & Blood density &$\text{g/cm}^3$ & 1.057 \cite{riches1973blood} \\
 $\mu_L$ & Viscosity & $\text{g/cm} \cdot \text{s}$&3.2 \cite{pries1992blood} \\
$\mu_S(r_0)$ & Viscosity module & $\text{g/cm} \cdot \text{s}$& see equation (\ref{eq:mu})\\
  $\nu$ & Kinematic viscosity &cm$^2$/s & $\mu/\rho$\\
 $\delta$ & Boundary layer thickness & cm &   $\sqrt{{\nu T}/{2 \pi}}$ \\
 $r_0$ & Given reference radius & cm & see Table \ref{tab:dimensions}\\
 $E$ & Young's modulus & mmHg & see equation (\ref{ehr0})\\
 $h$ & Wall thickness & cm & see equation (\ref{ehr0})\\
 $r_\text{min}$& Smallest vessel radius & mm & 0.001\\
 $l_{rr}$ & Length to radius ratio & Dimensionless & 15.75 (arteries);\\
\rowcolor{black!15}& & & 14.54 (veins) \cite{huang1996morphometry}\\
\rowcolor{white!50} $(\alpha,\beta$)& Structured tree scaling factors & Dimensionless & (0.84,0.67) \cite{chambers2020morphometry}\\
\midrule 
\textbf{Variables} & & &\\
\midrule
  $A(x,t)$ & Instantaneous cross-sectional area & cm$^2$ & computed\\
$q(x,t)$ & Volumetric flow rate & mL/s & computed\\
$p(x,t)$ & Blood pressure & mmHg & computed\\
 $r(x,t)$ & Instantaneous vessel radius & cm & computed \\
 $\tau_{w}$ &Wall shear stress & g/cm$\cdot$s$^2$ & computed\\
\midrule 
\end{tabular}
\end{center}
\end{table}

\section{Methods} \label{sec:methods}

\subsection{Network Geometry}

\begin{figure}[ht!]
  \centering
  \includegraphics[width=4in]{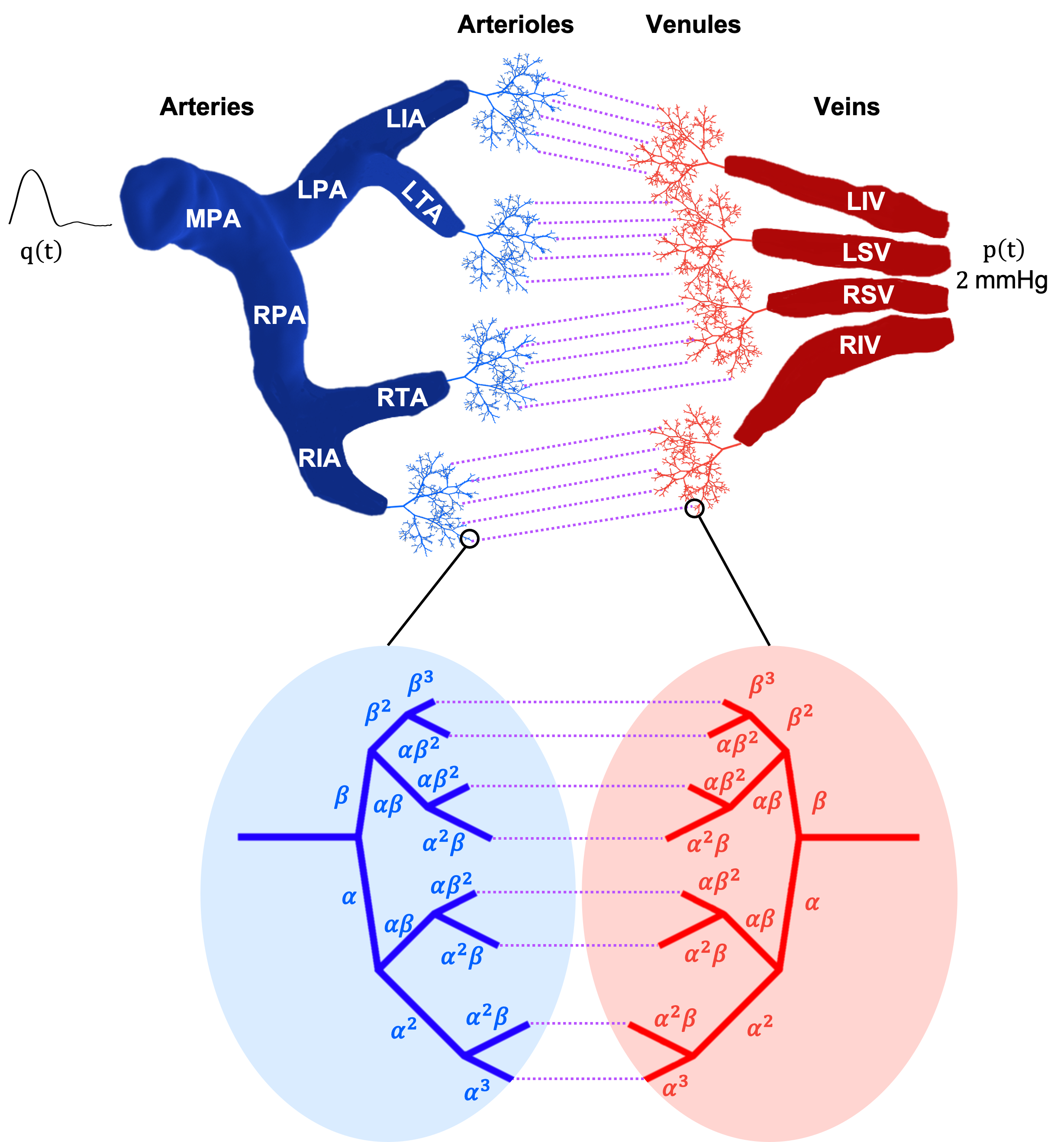}
  \caption{Schematic of the pulmonary circulation arranged in order of large arteries, arterioles, venules, and large veins. The dotted purple line represents the capillary beds which are not included in our model. The large arteries and veins are modeled explicitly using a CT image from a healthy human adult and the small vessels are modeled using a structured tree. In the structured tree, each small artery has a mirroring vein. The asymmetry of the structured tree is determined by the scaling factors, $\alpha$ and $\beta$.We impose an inflow profile at the MPA as well as a left atrial pressure of 2 mmHg at the pulmonary veins.}
\label{fig:modeldesign}
\end{figure}

Vessel dimensions and network connectivity were extracted from computed tomography (CT) chest images from a healthy 67 year old female volunteer\footnote{available from open-source Simvascular platform \url{http://simvascular.github.io/clinicalCase4.html}}. The geometric reconstruction of the first three generations of arteries and first generation of veins were generated using the open-source segmentation software 3DSlicer\footnote{\url{https://www.slicer.org/}} (see Figure \ref{fig:modeldesign}). We then used the vascular modeling toolkit (VMTK)\cite{antiga2008image} to obtain cartesian loci tracing of each vessel's center-lines within the network and measure associated radii. The VMTK point-cloud enabled the network topology quantification, in the form of connectivity matrices, and vessel's lengths computation by summing over the point-wise $L_2$ norms between successive center-line coordinates. Table \ref{tab:dimensions} contains the lengths and radii for the network of interest. To satisfy the topological equivalence of the connected arterial-venous structured trees, we set the same distal radii for the large terminal arteries (RIA, RTA, LIA, and LTA) and their mirroring veins (RSV, RIV, LSV, and LIV) \cite{qureshi2014numerical}.

Asymmetrically binary structured trees \cite{olufsen2000numerical} (shown in Figure \ref{fig:modeldesign}) are generated using physiology based scaling laws. At each junction, the daughter vessel radii and lengths are scaled by factors $\alpha$ and $\beta$ to the parent vessel, i.e.,
\[
     r_{d_1}=\alpha r_p,\quad r_{d_2}=\beta r_{p},\quad L_{(d_1,d_2)} = l_{rr}r_{(d_1,d_2)},
\]
where $\alpha,\beta<1$ and $\alpha+\beta\ge1$. Scaling parameters, along with a prescribed minimum radius $r_{m}$, determine the number of generation and vascular bed density encompassing the small arteries, arterioles, veins, and venules. Values for $\alpha$ and $\beta$ are taken from the study by Chambers et. al \cite{chambers2020morphometry}, estimating $\alpha$ and $\beta$ in vessels off the principal pathway. 

\begin{table}[ht!]
\caption{Dimensions of the large pulmonary vessels.}
    \label{tab:dimensions}
    \footnotesize
    \centering
\rowcolors{2}{black!15}{white!50}
\begin{center}
\begin{tabular}{ c c c c }    \toprule
&\textbf{Vessel Name}  & \textbf{Length (cm)} & \textbf{Radius (cm)} \\\midrule
& LSV & 2.42
& 0.70 \\ 
& LIV & 1.22 &
0.50
\\ 
& RSV & 4.74 &
0.80
\\ 
& RIV & 0.48 &
0.60
 \\ 
\midrule
 & LTA &2.01 &
0.70
 \\ 
& LIA & 2.45 &
0.50 \\ 
& RTA & 1.90 &
0.80
 \\ 
& RIA & 2.25 &
0.60
 \\
& LPA & 
6.24 &
1.19\\
& RPA & 
5.58 &
1.23\\ 
& MPA & 3.58 &
1.27 \\
\hline
\end{tabular}
 \end{center}
\begin{flushleft}
 \footnotesize{Main pulmonary artery (MPA), right (RPA) and left (LPA) pulmonary arteries, right (RIA) and left (LIA) interlobular arteries, right (RTA) and left (LTA) trunk arteries, right (RSV) and left (LSV) superior veins, and right (RIV) and left (LIV) inferior veins.}
 \end{flushleft}
    \end{table} 
\label{sec:data}

\subsection{Mathematical Model}\label{sec:model}

This study uses a 1D fluid dynamics network model, previously developed by Qureshi et al. \cite{qureshi2014numerical}, to compute blood flow, blood pressure, and area deformation throughout the pulmonary network (see Figure~\ref{fig:modeldesign}). This model delineates the hemodynamic domain into two sub-domains: a) a large vessel domain consisting of main arteries and veins modeled explicitly using dimensions from a healthy adult female extracted from the CT image, and b) a small vessel domain, composed of pre-capillary arterioles and venules generated using the two-sided morphology-based binary structured tree model. Mathematically, the large vessels belong to a nonlinear computational domain, requiring a numerical solution of the governing fluid equations, whereas the small vessels constitute a linearized analytical domain that combines exact solutions of governing fluid equations with a recursive algorithm. Below we describe the fluid dynamics modeling, resulting equations for the large and small vessels, as well as boundary conditions needed to predict the hemodynamic quantities of interest for this study.

\subsubsection{Large Vessel Fluid Dynamics} 
\label{sec:largevessels}

In large vessels, inertia dominates the blood flow dynamics; thus, pressure-flow relations can be determined from solving the Navier-Stokes equations in one dimension. Similar to our previous studies \cite{olufsen2000numerical,qureshi2014numerical} we compute blood pressure $p$ (mmHg), flow $q$ (mL/s), and dynamic cross-sectional area $A$ (cm$^2$) throughout the network. Each vessel is modeled as an axisymmetric tube with a circular cross-section and impermeable wall. We assume that the blood is Newtonian and the flow is incompressible, irrotational, and laminar. We account for the blood-wall interaction via the no-slip condition, ensuring that the blood flow velocity at the inner wall surface is same as the velocity of transverse wall motion. We further assume a Stokes boundary layer for axial velocity given by
\begin{align}
u_x(r,x,t) = \begin{cases}
\bar{u}, \ \ \ &r < R-\delta\\
\frac{\bar{u}(R-r)}{\delta}, \ \ \ &R - \delta < r \leq R
\end{cases}\label{eq:stokes}
\end{align}
where $\bar{u}_x$ is the averaged velocity outside of the boundary layer of thickness $\delta= \sqrt{\frac{\nu T}{2 \pi}}$ (cm) and $T$ (s) is cardiac cycle length. The Stokes layer provides a flat axial velocity profile characterizing the inertia-dominant flow in the large vessels. We obtain the conservation of mass and balance of momentum by averaging the 1D Navier-Stokes equations, giving
\begin{equation}
    \pard{A}{t}+\pard{q}{x}=0,\quad\qquad \pard{q}{t}+\pard{}{x}\left(\frac{q^2}{A}\right)+\frac{A}{\rho}\pard{p}{x}=-{\cal F}\frac{q}{A},
\end{equation}
where $x$ (cm) and $t$ (s) are spatial and temporal coordinates, respectively, ${{\cal F} = 2\pi \nu/\delta}$ is a frictional term, and $\nu=\mu/\rho$ (cm$^2$/s) is the kinematic viscosity. To close the system of equations, we assume a linear elastic relationship between pressure and cross-sectional area of the form
\begin{align}
    p(x,t)-p_0 = \frac{4}{3}
    \frac{Eh}{r_0}\left(\sqrt{\frac{A}{A_0}}-1\right),
\end{align}
where $E$ (mmHg) is Young's modulus, $h$ (cm) is vessel wall thickness, $p_0$ is a reference pressure (mmHg), and $A_0$ (cm$^2$) is the corresponding cross-sectional area. To account for the fact that smaller vessels are stiffer, we model
\begin{align}
    \frac{Eh}{r_0} = k_1 \exp(-k_2 r_0)+k_3  \label{ehr0}
\end{align}
for the small vessels. For small arteries, we assume that $k_1 = 2.6 \times 10^5 $ (g/s$^2\cdot$ cm),  $k_2=-14$ cm$^{-1}$, and $k_3 = 10 \times 10^4$  (g/s$^2\cdot$ cm). For the small veins, the respective stiffness coefficients are doubled from that of the veins. For the large vessels, we assume a constant stiffness of $6 \times 10^5$  g/s$^2\cdot$ cm.

Since our system of equations is hyperbolic, we specify boundary conditions at the inlet and outlet of each vessel. At the inlet to the pulmonary circulation, i.e., the main pulmonary artery (MPA), we specify a flow waveform extracted from data (Figure \ref{fig:modeldesign}). At network junctions we enforce flow conservation and pressure continuity
\begin{align}
    q_p(l_p,t)&=q_{d_1}(0,t)+q_{d_2}(0,t)\\
    p_p(l_p,t)&=p_{d_1}(0,t)=p_{d_2}(0,t),
\end{align}
where the subscripts $p, d_1, d_2$ specify parent and daughter vessels and $l_p$ specifies the length of the parent vessel. The outflow from the large terminal arteries and the inflow into the corresponding large terminal veins are obtained by coupling these vessels in series to two-sided arterial and venous structured trees. At the outflow of the large veins, a constant pressure of 2 mmHg is specified, corresponding to the mean left atrial pressure in a healthy adult.

\subsubsection{Small Vessel Fluid Dynamics} 

\begin{figure}
    \centering
    \includegraphics[width=4 in]{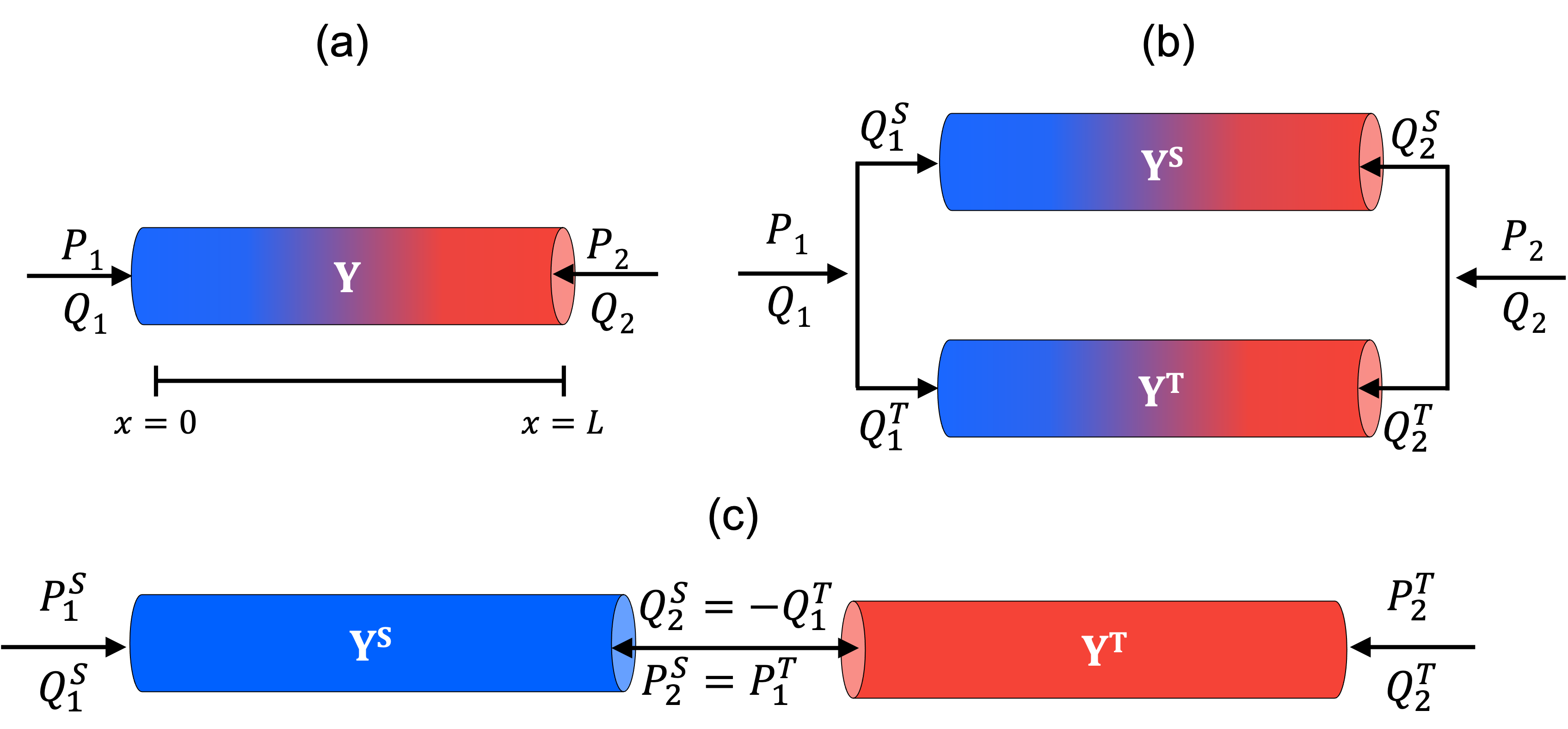}
    \caption{Relationship between flow and pressure via admittance $\mathbf{Y}$, $\mathbf{Y}^{\|}$ and $\mathbf{Y}^{\Leftrightarrow}$ for a single vessel (a), vessels connected in parallel (b), and vessels connected in series (c). Here, the color blue represents deoxygenated arteries and red represents oxygen-rich veins.}
    \label{fig:admittance}
\end{figure}

In the small vessels, viscous effects are dominant, so we can ignore the nonlinear inertial terms. As described in detail in \cite{olufsen2000numerical,qureshi2014numerical}, the momentum equation in the axial direction is given by
\begin{align}
  \pard{u_x}{t}+\frac{1}{\rho}\pard{p}{x} = \frac{\nu}{r}\pard{}{r}\left(r\pard{u_x}{r}\right),
 \label{eqn:floweq}
\end{align}
and the continuity equation is
\begin{align}
C \pard{p}{t}+\pard{q}{x} = 0 \label{eq:eq}, \ \ \ 
C=\pard{A}{p}=\frac{3A_0r_0}{2Eh}\left(1-\frac{3pr_0}{4Eh}\right)^{-3}\approx\frac{3A_0r_0}{2Eh}.
\end{align}
Assuming periodicity, flow and pressure can be written as $q(x,t)=Q(x)e^{i\omega t}$ and $p(x,t)=P(x)e^{i\omega t}$. Using this decomposition we can rewrite the momentum and continuity equations as 
\begin{align}
& i\omega Q + \frac{A_0(1-F_J)}{\rho}\frac{\partial P}{\partial x} = 0,  \ \ \ F_J = 2J_1(w_0)/w_0 J_0(w_0) \label{eq:eq1}\\
&i\omega CP+\pard{Q}{x}=0, \label{eq:eq2}
\end{align}
where $J_i(w_0), i=0,1$ are the first and second order Bessel functions, and $w_0=i^3 r_0^2\omega/\mu$ is the Womersley number. Since the effects of blood viscosity become more significant as the vessel size decreases, as suggested by Pries et. al \cite{pries1992blood}, we compute viscosity as a function of the unstressed vessel radius $r_0$.
\begin{align}
    \mu(r_0) &= \left[1+(\mu_{0.45}-1)\left(\frac{2r_0}{2r_0-1.1}\right)^2\right]\left(\frac{2r_0}{2r_0-1.1}\right)^2  \label{eq:mu} \\
  \mu_{0.45}(r_0)&= 6e^{-0.17r_0} +3.2-2.44e^{-0.12r_0^{0.645}},
\end{align}
where $\mu_{0.45}(r_0)$ is the relative viscosity at a hematocrit level of 0.45 \cite{pries1992blood}.
Differentiating equation (\ref{eq:eq2}) with respect to $x$ and substituting the result into equation (\ref{eq:eq1}) gives a wave equation of the form
\begin{align}
\frac{\omega^2}{c^2}Q+\pard{^2Q}{x^2}=0 \ \ \text{ or } \ \ 
\frac{\omega^2}{c^2}P + \pard{^2P}{x^2}=0 \end{align}
where $c=\sqrt{A_0(1-F_J)/\rho C}$ is the wave propagation velocity.  The solution to these equations is given by
\begin{align}
    Q &= a\cos\left(\frac{\omega x}{c}\right) + b\sin\left(\frac{\omega x}{c}\right) \label{eqn:q}\\
    P &= i {g_\omega}^{-1}\left(-a\sin\left(\frac{\omega x}{c}\right)+\cos\left(\frac{\omega x}{c}\right) \right), \  g_\omega =\sqrt{\frac{CA_0(1-F_J)}{\rho}}.
    \label{eqn:pxw}
\end{align}
As described in detail by Qureshi et al. \cite{qureshi2014numerical}, for the arterial and venous networks we define a relation between pressure and flow at both ends of each vessel, and  derive a boundary condition that matches pressure and flow at the terminal large arteries and veins. To do so, we compute the admittance of each structured tree, relating pressure and flow at the outlet of each large terminal artery to pressure and flow at the inlet to the corresponding large terminal vein. At each junction, we set up an admittance matrix satisfying continuity of pressure and conservation of volume flux. The total admittance is calculated by joining ``junction" admittances in series and parallel as illustrated in Figure \ref{fig:admittance}.

Given the proximal and distal flow $Q_{1}=Q(0, \omega)$ and $Q_{2}=Q(L, \omega)$, the admittance matrix is obtained by relating the flow and pressure at the proximal, $x=0,$ and distal, $x=L,$ ends of a vessel of radius $r_0$ yielding
\begin{align}
\left(\begin{array}{l}Q_{1} \\ Q_{2}\end{array}\right)=\frac{i g_{\omega}}{S_{L}}\left(\begin{array}{cc}-C_{L} & 1 \\ 1 & -C_{L}\end{array}\right)\left(\begin{array}{l}P_{1} \\ P_{2}\end{array}\right)
\end{align}
where $C_{L} \equiv \cos (\omega L / c), S_{L} \equiv \sin (\omega L / c),$ and
\begin{align}
\mathbf{Y}(\omega)=\frac{i g_{\omega}}{S_{L}}\left(\begin{array}{cc}-C_{L} & 1 \\ 1 & -C_{L}\end{array}\right)
\end{align}
is the admittance matrix for any one artery or vein when $\omega \neq 0.$ For $\omega=0,$ we have
\begin{align}
\left(\begin{array}{l}Q_{1} \\ Q_{2}\end{array}\right)=\frac{\pi r^{4}}{8 \mu L}\left(\begin{array}{rr}1 & -1 \\ -1 & 1\end{array}\right)\left(\begin{array}{l}P_{1} \\ P_{2}\end{array}\right).
\end{align}
and therefore
\begin{align}\mathbf{Y}(0)=\frac{\pi r^{4}}{8 \mu L}\left(\begin{array}{rr}1 & -1 \\ -1 & 1\end{array}\right).
\end{align}
For two vessels ($S$ and $T$) in parallel, continuity of pressure between the inlet and outlet  and conservation of volume flux across the junction gives
\begin{align}
    \left(\begin{array}{l}Q_{1} \\ Q_{2}\end{array}\right)=\mathbf{Y}^{\|}\left(\begin{array}{l}P_{1} \\ P_{2}\end{array}\right),
    \end{align}
where $Q_{1}$ and $Q_{2}$ denote the vessel inflow and outflow, $P_1$ and $P_2$ are the corresponding inlet and outlet pressure, and $\mathbf{Y}^{\|}=\mathbf{Y}^{S}+\mathbf{Y}^{T}$.

Similarly, for two vessels ($S$ and $T$) connected in series flow and pressure are related by
\begin{align}
Q_{k}^{i}=\sum_{l=1}^{2} Y_{k l}^{i} P_{l}^{i}
\end{align}
where $i=S, T$ and $k=1,2$ with $Y_{k l}$ being the components of the $2 \times 2$ admittance matrix. Assuming that $P=P_{2}^{S}=P_{1}^{T}$ and $Q_{2}^{S}=-Q_{1}^{T}$ at the junction of two vessels in series, the system for the vessels is represented by
\begin{align}
\left(\begin{array}{l}
Q_{1}^{S} \\
Q_{2}^{T}
\end{array}\right)=\mathbf{Y}^{\Leftrightarrow}\left(\begin{array}{l}
P_{1}^{S} \\
P_{2}^{T}
\end{array}\right)
\end{align}
where
\begin{align}
\mathbf{Y}^{\Leftrightarrow}=& \frac{1}{Y_{22}^{S}+Y_{11}^{T}}  \times\left(\begin{array}{cc}\operatorname{det}\left(\mathbf{Y}^{S}\right)+Y_{11}^{S} Y_{11}^{T} & -Y_{12}^{S} Y_{12}^{T} \\
-Y_{21}^{S} Y_{21}^{T} & \operatorname{det}\left(\mathbf{Y}^{T}\right)+Y_{22}^{S} Y_{22}^{T}
\end{array}\right)
\end{align}
is the admittance matrix. The symbol $\Leftrightarrow$ denotes that the vessels are joined in series.

To compute pressure and flow in the arterioles and venules, we first prescribe the pressure from the terminal large arteries or veins at the start of the structured tree, i.e. $P^A(0,\omega)$ and $P^V(0,\omega)$. The corresponding flows are calculated using the grand admittance, 

\begin{align*}
    Q^A(0,\omega) = Y_{11}P^A(0,\omega) + Y_{12}P^V(0,\omega), \\
    Q^V(0,\omega) = Y_{21}P^A(0,\omega) + Y_{22}P^V(0,\omega).
\end{align*}

The small vessel pressure and flow at $x=L$ are determined by

\begin{align}
    P^j(L,\omega) = P^j(0,\omega)C_L - \frac{iS_L}{g_\omega}Q^j(0,\omega), \\
    Q^j(L,\omega) = \frac{i g_\omega}{S_L}\left(P^j(0,\omega)-C_L P^j(L,\omega)\right)
\end{align}
for $j=A,V$ and $\omega\neq0$, whereas the zeroth frequency solutions are
\begin{align}
    P^j(L,\omega) = P^j(0,\omega)- \frac{8 \mu L}{\pi r^4}Q^j(0,\omega), \\
    Q^j(L,\omega) = \frac{\pi r^4}{8 \mu L}\left(P^j(0,\omega)-P^j(L,\omega)\right).
\end{align}

\subsubsection{Wall Shear Stress (WSS)}

\paragraph{Large Vessels.} The shear stress, $\tau_w$, that the fluid exerts on the vessel wall can be computed from the Stokes boundary layer Equation (\ref{eq:stokes}), giving
\begin{align}
\tau_w&= - \mu\pard{u}{r}\\
    &= \begin{cases}
    0, \ \ \ &r < R-\delta\\
    \dfrac{\mu \bar{u}}{\delta},\ \ \  &R - \delta < r \leq R.
    \end{cases}
\end{align} 

\paragraph{Small Vessels.} To compute the shear stress in the small vessels, we assume (consistent with the pumping action of the heart) that the driving pressure is oscillatory in time. 

\par We assume that this driving force consists of a constant part that does not vary in time and an oscillatory part that moves the fluid back and forth over each cycle \cite{zamir2002physics}. We refer to the combination of the two components as pulsatile flow. The steady and oscillatory components of pressure and axial velocity are denoted by
\begin{align}
    p(x,t) = p_s(x) + p_\phi(x,t) \ \ \ \text{and} \ \ \
    u(r,t) = u_s(r) + u_\phi(r,t).
\end{align}
 Because of the independence of oscillatory and steady flow, the relationship between the pressure gradients is given by
 \begin{align}
 k(t) &= k_s +k_\phi(t),
 \end{align}
where $k_\phi(t) = k_s e^{i\omega t}$ and $u_\phi(r,t) = U_\phi(r) e^{i\omega t}.$ Substituting these expressions into Equation \ref{eqn:floweq} gives
 \begin{align}
    \pard{^2U_\phi}{r^2}+\frac1r \pard{U_\phi}{r} -\frac{i \Omega^2 }{a^2} U_\phi(r) = \frac{k_s}{\mu}, \label{eqn:solve}
 \end{align}
 where $\Omega = R\sqrt{\frac{\rho \omega}{\mu}}$, $R$ is the tube radius, and $\mu$ is radius dependent. The solution is given by 
 \begin{align}
     U_\phi = \frac{ik_s R^2 }{\mu \Omega^2}\left(1- \frac{J_0(\zeta)}{J_0(\Lambda)}\right), 
 \end{align}
where $\zeta(r) = \Lambda \frac{r}{R} \text{ and } \Lambda = \left(\frac{i-1}{\sqrt{2}}\right)\Omega$.
\label{sec:smallvessels}
To find the oscillatory volumetric flow rate $q_\phi$ through a tube, we integrate the oscillatory velocity over a cross-section of the tube 
\begin{align*}
q_\phi(t) &=\int_0^a 2 \pi ru_\phi(r,t)\ dr\\
&= \frac{i\pi k_s R^4}{\mu \Omega^2}\left(1-\frac{2}{\Lambda}\frac{J_1(\Lambda)}{J_0(\Lambda)}\right)e^{i\omega t}.
\end{align*}
As the fluid moves back and forth due to an oscillatory pressure gradient, the shear stress exerted by the fluid on the tube wall is given by
\begin{align}
    \tau_\phi(t) &= - \mu \left(\pard{u_\phi(r,t)}{t}\right)_{r=R}\\
    &= -\frac{ik_sR^2}{\Omega^2}\left[\frac{d}{dr}\left(1-\frac{J_0(\zeta)}{J_0(\Lambda)}\right) \right]_{r=R}e^{i\omega t}\\
    &= -\frac{ik_sR^2}{\Omega^2} \left[\frac{d}{d\zeta}\left(1-\frac{J_0(\zeta)}{J_0(\Lambda)}\right) \right]_{\zeta=\Lambda}\frac{\Lambda}{R}e^{i\omega t}\\
    &= -\frac{k_sR}{\Lambda} \left(\frac{J_1(\Lambda)}{J_0(\Lambda)}\right) e^{i\omega t}\\
   &= \frac{\Lambda}{\pi R^3}\frac{J_1(\Lambda)}{J_0(\Lambda)-\frac2\Lambda J_1(\Lambda)}q_\phi.
\end{align}
This expression is complex with its real part representing shear stress at the tube wall when the driving pressure gradient varies as $\cos(\omega t)$ and its imaginary part representing shear stress at the wall when the driving pressure varies as $\sin(\omega t).$ In both cases, the shear stress is scaled with the shear stress in Poiseuille flow \cite{zamir2002physics}. 

To compare our results to 3D studies we compute one-dimensional analogies to TAWSS and OSI. We compute TAWSS by finding
\begin{align*}
    \text{TAWSS} = \frac{1}{T}
 \int_0^T \tau_w \ dt.
 \end{align*}
 Since we take the period, $T$, to be $1$ (Table \ref{tab:defn}), this is simply the average WSS. To find OSI, we divide the TAWSS by the change in amplitude of WSS over time. This allows us to find the mean WSS over the pulsation, which in turn exhibits the magnitude of oscillation of the WSS. 
 
 We also quantify CS in order to develop an understanding of how diameter and circumferential stretch are related. To do this, we calculate
 \begin{align*}
     \text{CS} = \frac{A_\text{max}-A_\text{min}}{A_\text{min}}\times 100
 \end{align*}
 to find the percentage of change.
 
\section{Simulations}

We use the two-step Richtmeyer Lax-Wendroff method to solve the model equations described in Section \ref{sec:methods} for the large arterial and venous networks presented in Table \ref{tab:dimensions}. Using data from Table \ref{tab:dimensions} and methods described for the structured tree model, we predict pressure, area, and flow in the large pulmonary arteries and veins. We also compute the WSS, TAWSS, and OSI for these large vessels using the Stokes boundary layer velocity profile. Using the linearized model equations to simulate the microvasculature, we predict pressure and flow along the arteriole and venule $\alpha$ and $\beta$ branches and observe how they vary as vessel size decreases. We also compute CS, WSS, TAWSS, and OSI in the microvasculature.

\section{Results}

\begin{figure}[ht!]
    \centering
    \includegraphics[width = 4 in]{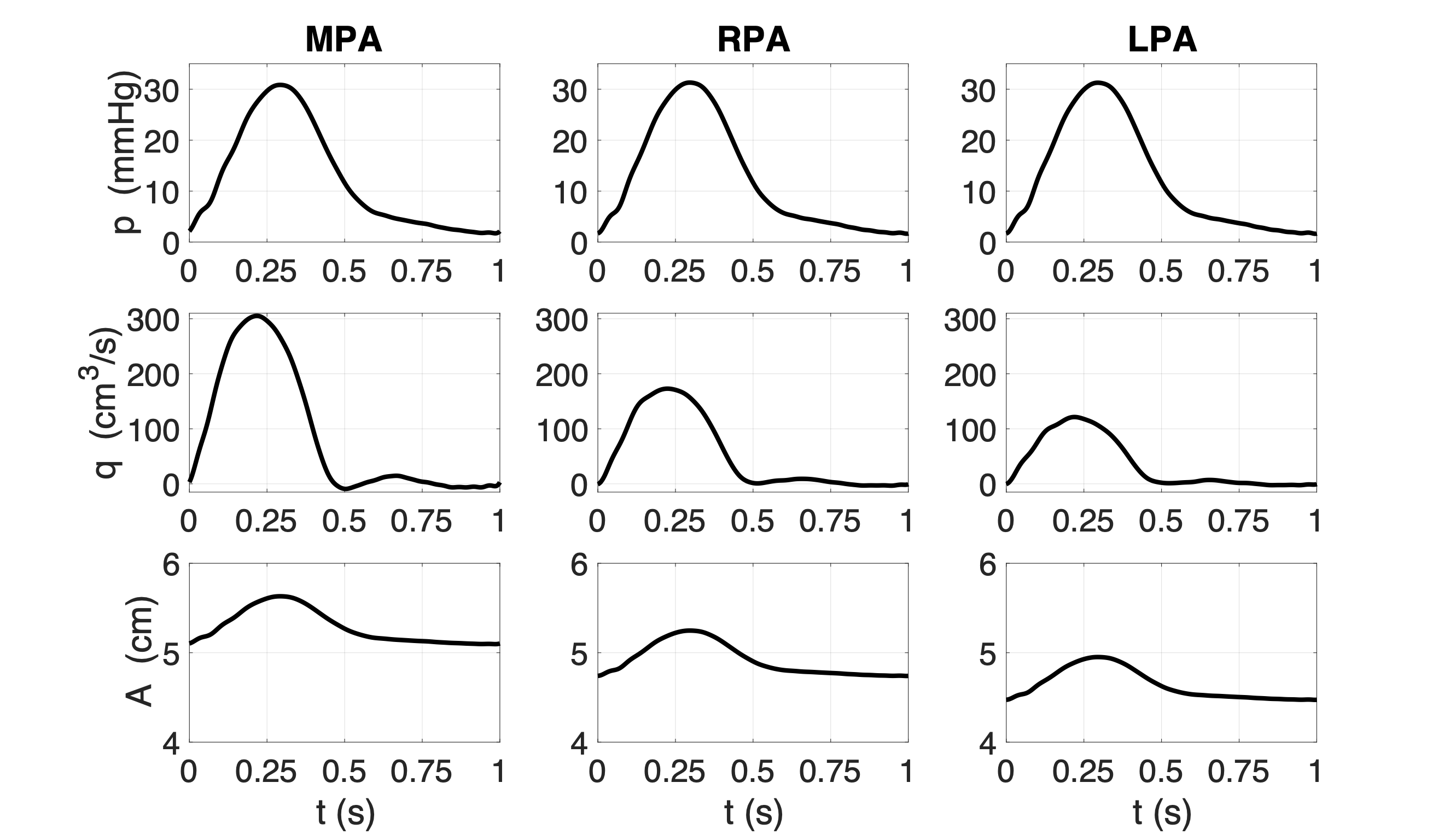}
    \caption{Predicted pressure (first row), flow (second row), and area (third row) at three locations along the large arteries, MPA (first column), RPA (second column), and LPA (third column), for  left atrial pressure of 2 mmHg and typical cardiac output}
    \label{fig:PQA_arteries}
\end{figure}

\begin{figure}[ht!]
    \centering
    \includegraphics[width=\textwidth]{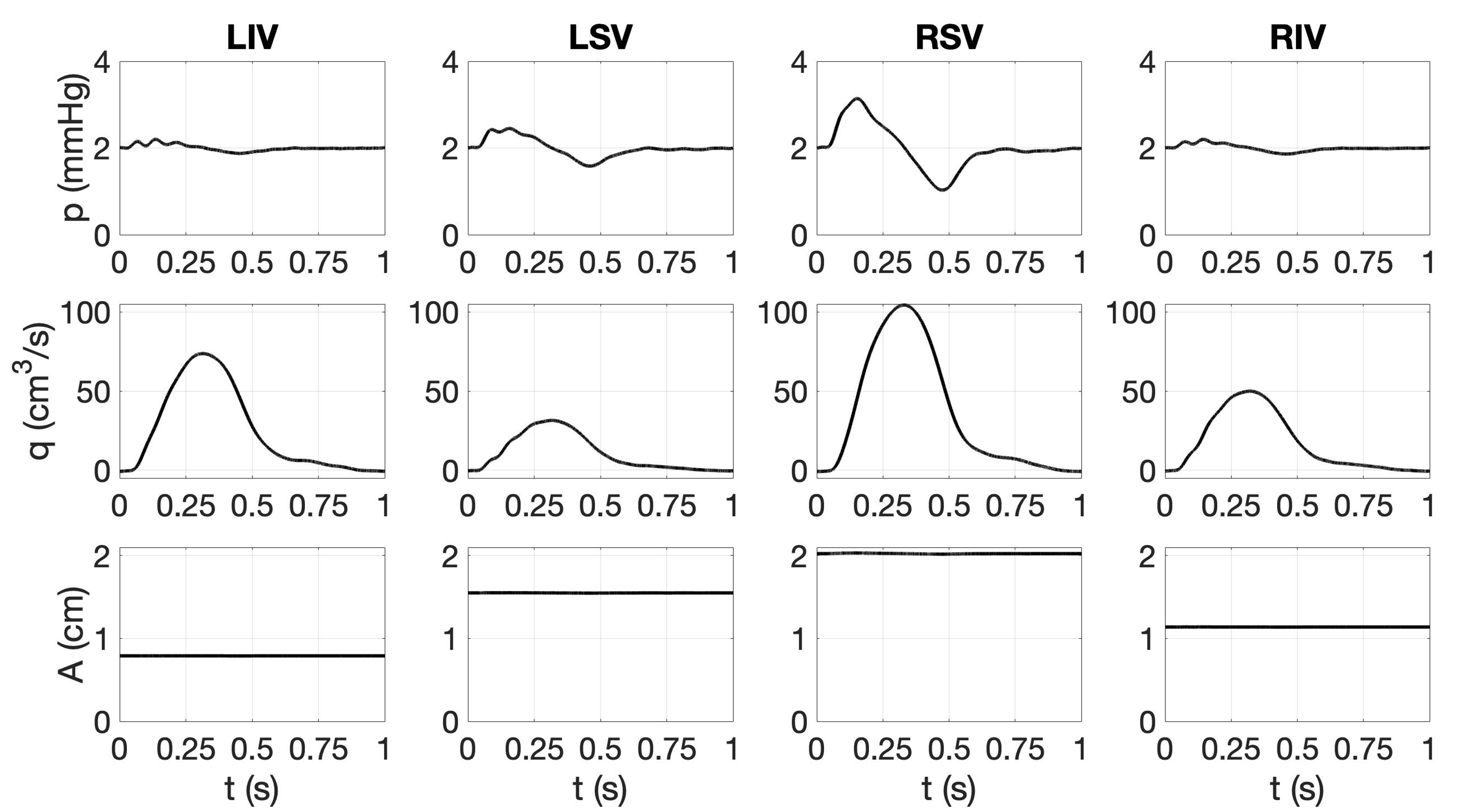}
    \caption{Predicted pressure (first row), flow (second row), and area (third row) at the four large pulmonary veins, LIV (first column), LSV (second column), RSV (third column), and RIV (fourth column) for left atrial pressure of 2 mmHg and typical cardiac output}
    \label{fig:PQA_veins}
\end{figure}

We show results for a healthy female adult, with network dimensions specified in Table \ref{tab:dimensions}. Vessel geometries and inflow into the MPA were obtained from measurements; other quantities including density and viscosity were determined from literature values (Table \ref{tab:defn}). Figure \ref{fig:PQA_arteries} shows the predicted flow, pressure, and area at the midpoint of the MPA, RPA, and LPA over one cardiac cycle. The pressure ranges from approximately 8 mmHg to 30 mmHg, and as expected, the mean flow decreases as it is distributed to the downstream vasculature. The pressure for large veins remains close to 2 mmHg with the LSV and RSV having the greatest pulse pressure over a cardiac cycle (Figure \ref{fig:PQA_veins}). With a cardiac output corresponding to a healthy adult, and a heartrate of 60 beats per minute, the maximum flow for the MPA, RPA, and LPA are 300 $\text{cm}^3/\text{s}$, 160 $\text{cm}^3/\text{s}$, and 100 $\text{cm}^3/\text{s}$, respectively. The volumetric flow for each of the large veins is less than that of the arteries (Figure \ref{fig:PQA_arteries},\ref{fig:PQA_veins}). 
The maximum flow for the LIV, LSV, RSV, and RIV are approximately $75$ $\text{cm}^3/\text{s}$,  40 $\text{cm}^3/\text{s}$, 100 $\text{cm}^3/\text{s}$, and 50 $\text{cm}^3/\text{s}$, respectively. 

\par We simulate the area over a cardiac cycle for the large arteries and veins. We observe that the MPA has the largest area, ranging from approximately 5 to 5.6 cm over a cardiac cycle. Area for RPA ranges from 4.7 to 5.2 cm and area for LPA ranges from 4.5 to 5 cm. The area for the large veins does not fluctuate as much as the large arteries throughout a cardiac cycle and is less than that of the arteries. This is due to the fact that veins are twice as stiff as arteries in our model.

\par
\begin{figure}[ht!]
    \begin{center}
    \includegraphics[width=\textwidth]{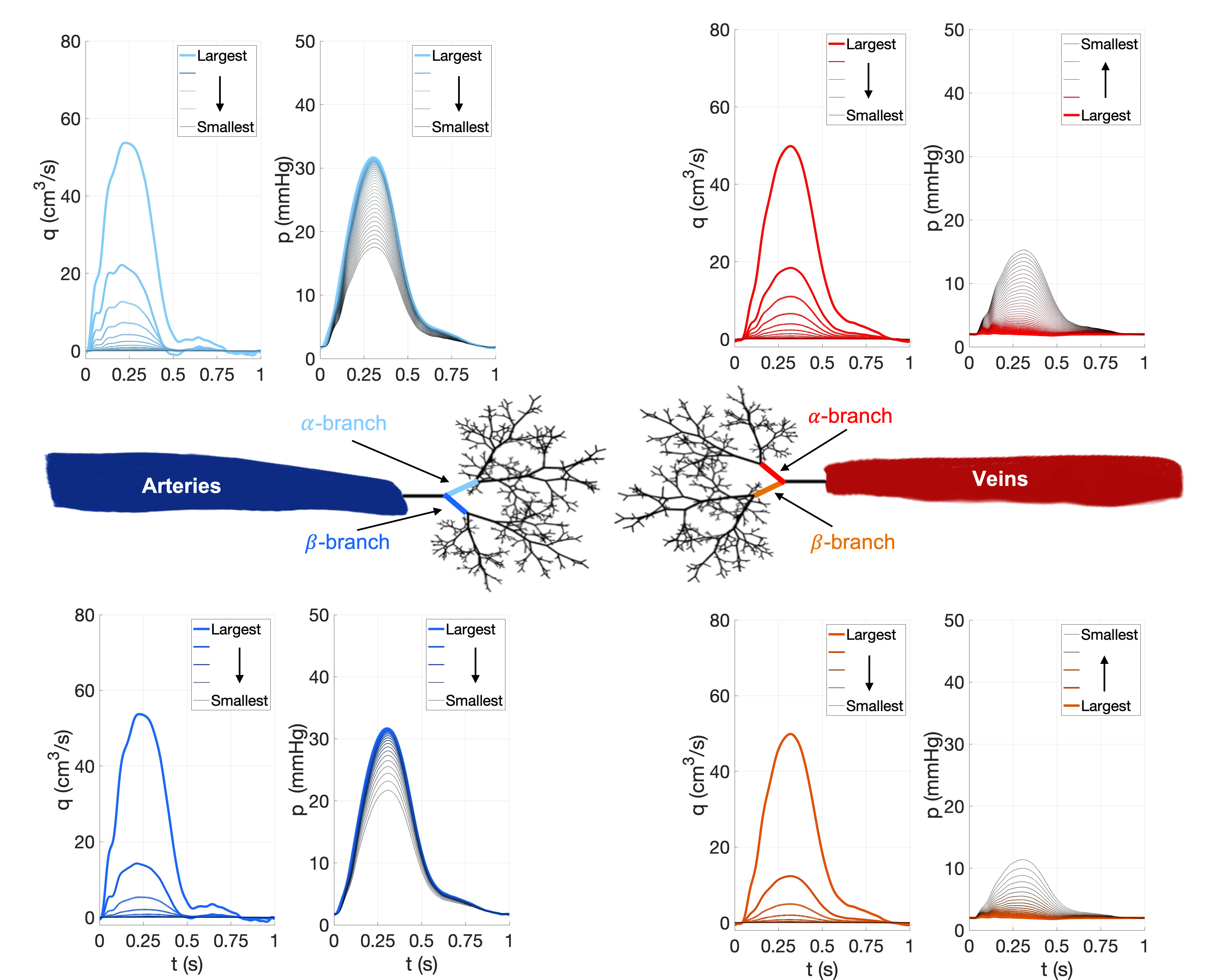}
 \end{center}
    \caption{Predicted flows and pressure for small arteries and veins along the $\alpha$ and $\beta$ branches of the structured tree. Arteries are shown in the first column and veins are shown in the second column with results for the $\alpha$ branch shown in the top row and results for the $\beta$ branch shown in the bottom row. The largest vessels are represented by colored thick lines and as we move down the structured tree, the smallest vessels are represented by thin black lines.}
    \label{fig:flowsmallart}
\end{figure}

 We also predict pressure and flow in the small arteries and veins along the $\alpha$ and $\beta$ branches (Figure \ref{fig:flowsmallart}). In the arterioles, the pressure decreases with decreasing vessel radius with a maximum pressure of about 30 mmHg and minimum pressure of about 2 mmHg. By contrast, in the venules, the smaller vessels have a larger pressure. Since we set $r_m$ to be $0.001$ mm, all vessels that have a smaller radius than this value are excluded from the model. This leads to a pressure discontinuity between the small arteries and veins as we do not include vessels with radii less than $0.001$ mm. In addition, due to the asymmetry of the structured tree model, the $\beta$ branch has fewer vessels than the $\alpha$ branch. This leads to a larger gap in pressure between arteries and veins in the $\beta$ branch. The minimum pressure in the small veins approaches 2 mmHg at the inlet to the left atrium. For both arterioles and venules, flow decreases with decreasing vessel radius.

\par
\begin{figure}[ht!]
    \centering
    \includegraphics[width=0.75\textwidth]{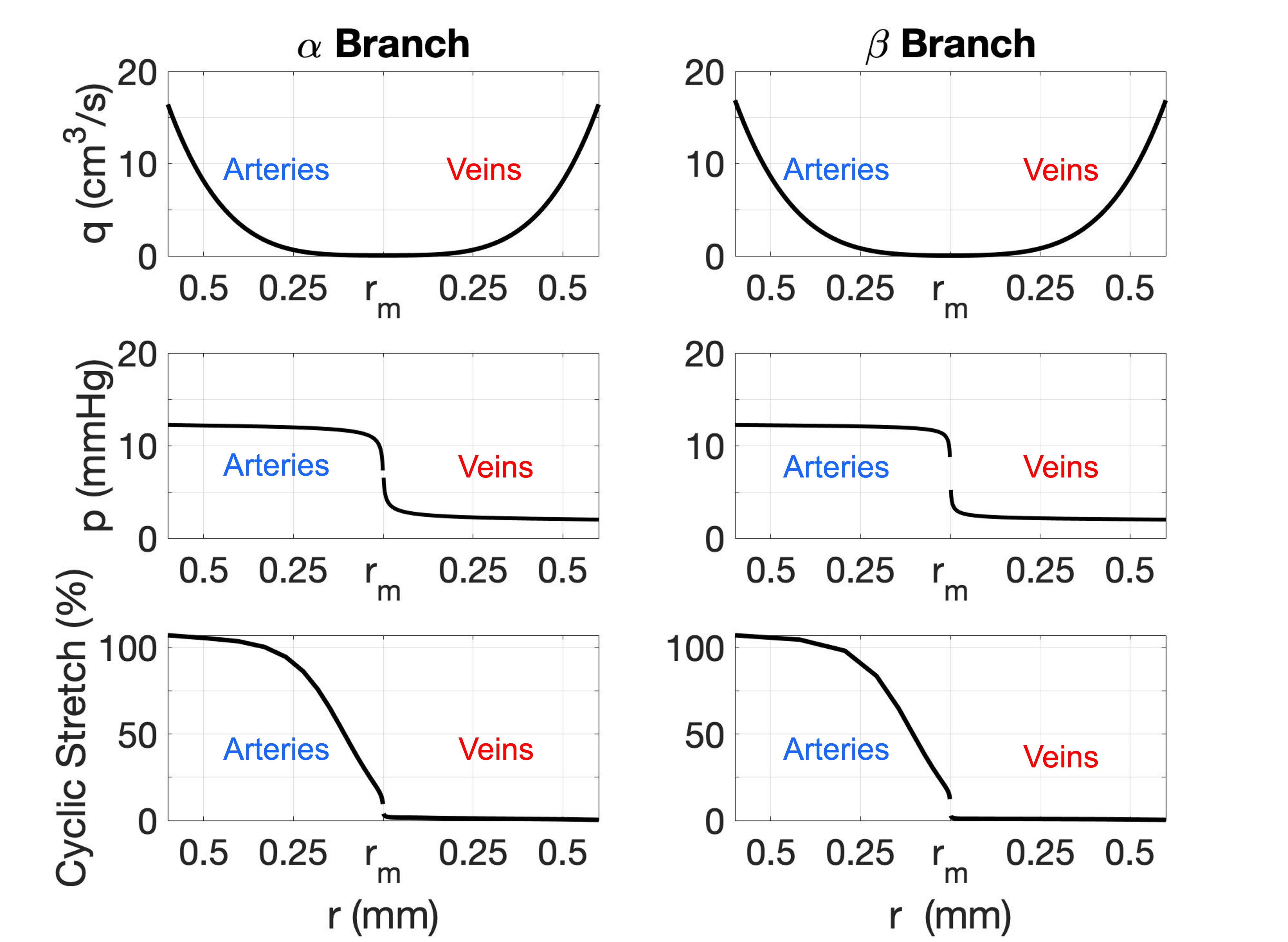}
    \caption{Mean flow, pressure, and cyclic stretch for the $\alpha$ branch (left) and $\beta$ branch (right) for small arteries and veins of the structured tree. The gap in the pressure plots are due to the minimum radius, $r_m$, that is set in the model. 
    There are fewer vessels in $\beta$ branch, which leads to a larger gap being shown.}
    \label{fig:meanflowpres}
\end{figure}

 CS is another physiological measure that can be used to understand endothelial function and we show predictions for CS in the arterioles and venules in Figure \ref{fig:meanflowpres}.  CS in the small arteries with radii between 0.5 mm and r$_\text{m}$ decreases from 100\% to approximately 10\%. The arterioles have a higher percentage of CS and CS decreases with decreasing vessel radius. For venules, CS remains around about 5\%, increasing slightly with decreasing vessel radii. 

\par
\begin{figure}[ht!]
    \centering
    \includegraphics[width=\linewidth]{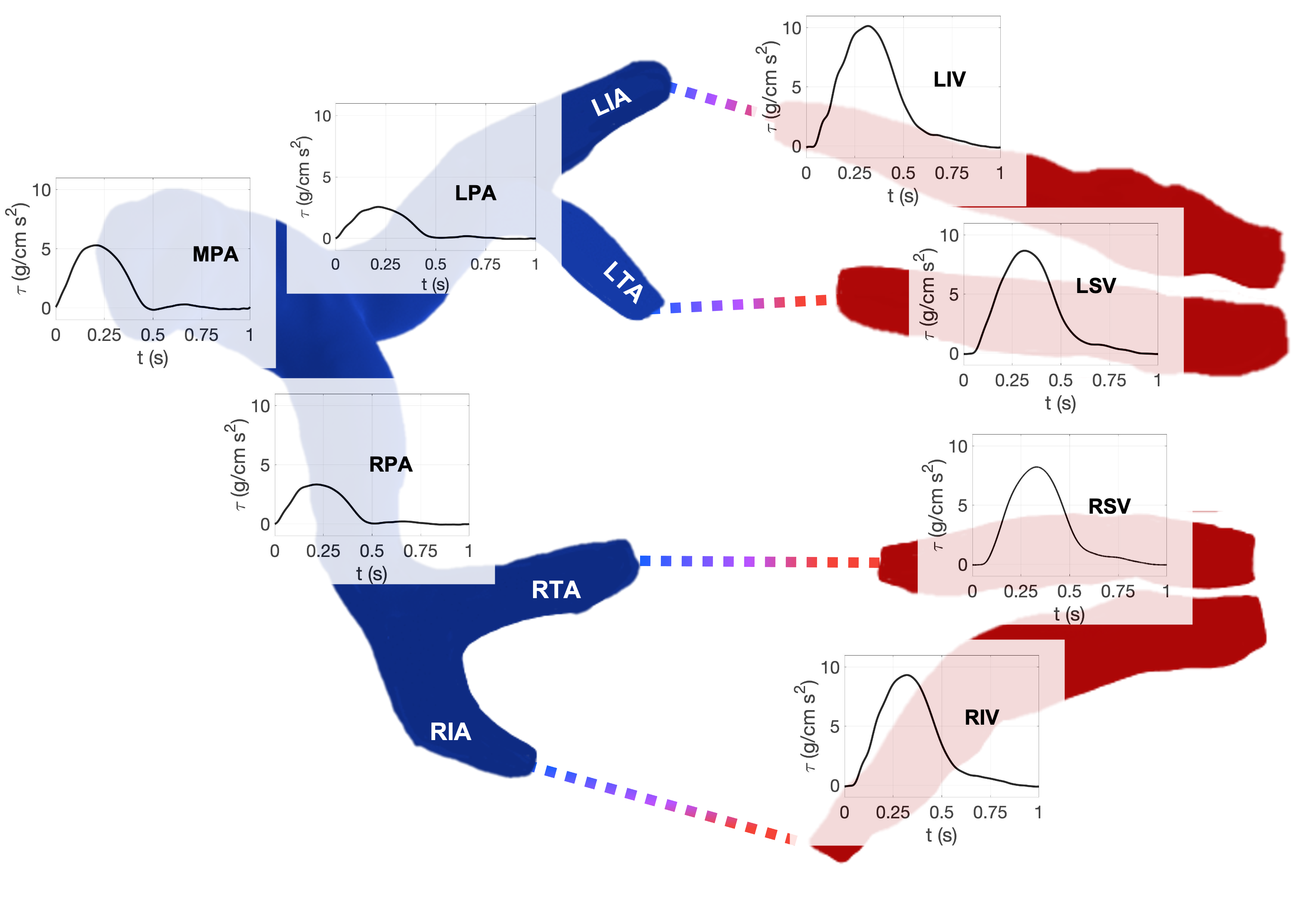}
    \caption{Shear stress using a Stokes boundary layer formation along three locations along the large arteries (MPA, RPA, LPA) and all four large veins for left arterial pressure of 2 mmHg and typical cardiac output.}
\label{fig:shearstress}
\end{figure}

\begin{figure}[ht!]
    \centering
    \includegraphics[width=0.85\textwidth]{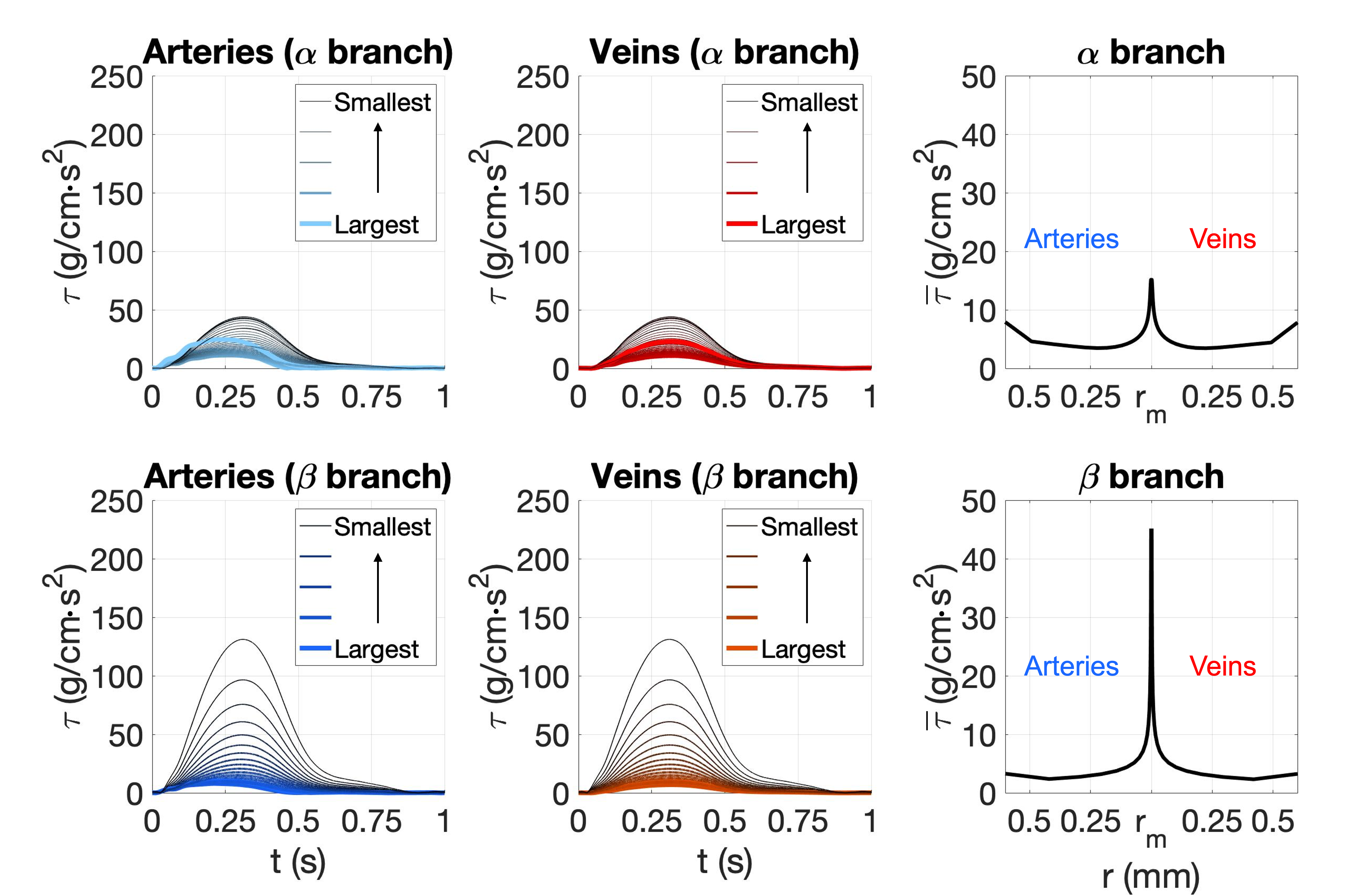}
    \caption{Predicted shear stress for small arteries and veins along the $\alpha$ and $\beta$ branches of the structured tree (first two columns). The largest vessels are shown in thick colored lines and line thickness and color decreases with vessels. Average shear stress for arteries and veins along the $\alpha$ and $\beta$ branches is plotted against radius (third column).}
    \label{fig:shearstresssmall}
\end{figure}

 We also compute time-varying shear stress in the large arteries and veins (Figure \ref{fig:shearstress}) and in the small arteries and veins (Figure \ref{fig:shearstresssmall}). For the large vessels, we compute the shear stress within the Stokes boundary layer at the midpoint of the main (MPA), right (RPA), and left (LPA) pulmonary arteries, and the four large veins (LIV, LSV, RSV, RIV). The MPA has the largest shear stress. In general, the veins have a higher shear stress than arteries. Shear stress for the veins ranges between 0 - 7 g/cm$^2$s, whereas the shear stress for the arteries ranges between 0 - 5 g/cm$^2$s. For the small vessels, shear stress is predicted along the $\alpha$ and $\beta$ branches to investigate the impact of the asymmetry of the structure tree. In both the small arteries and veins, smaller vessel radius results in higher values for shear stress. We also see that the $\beta$ branch has higher shear stress than the $\alpha$ branch. 
The $\beta$-branch has fewer vessels than the $\alpha$-branch due to the asymmetric set up of the structured tree, while still having similar total flow. This results in the $\beta$-branch having higher shear stress.

\par We also average the wall shear stress over time (TAWSS) and plot the results at the midpoint of each vessel. The TAWSS ranges from about 0.75 to 3 g/cm$^2$s. Vessels with smaller radii (RTA, RIA, LTA, LIA) have a higher shear stress. We also observe that the MPA has a higher TAWSS than the RPA and LPA, despite the latter having smaller radii. This may be because the MPA has more flow within a cardiac cycle than the RPA and LPA. For the small vessels, we plot the TAWSS with respect to vessel radius and observe that vessels with the smallest radii have the largest TAWSS. This follows from the fact that vessels with smaller radii typically have larger WSS.

\par For both the large and small arteries, the OSI ranges from approximately 0.28 to 0.34. The small veins have a nearly constant OSI at approximately 0.34. For the large vessels, OSI increases throughout the tree with the MPA having the smallest OSI. For the small vessels, we observe that the venules have consistently higher OSI than the arterioles demonstrating that venules are more oscillatory than the arterioles.

\label{sec:results}


\begin{figure}[ht!]
    \centering
    \includegraphics[width=4in]{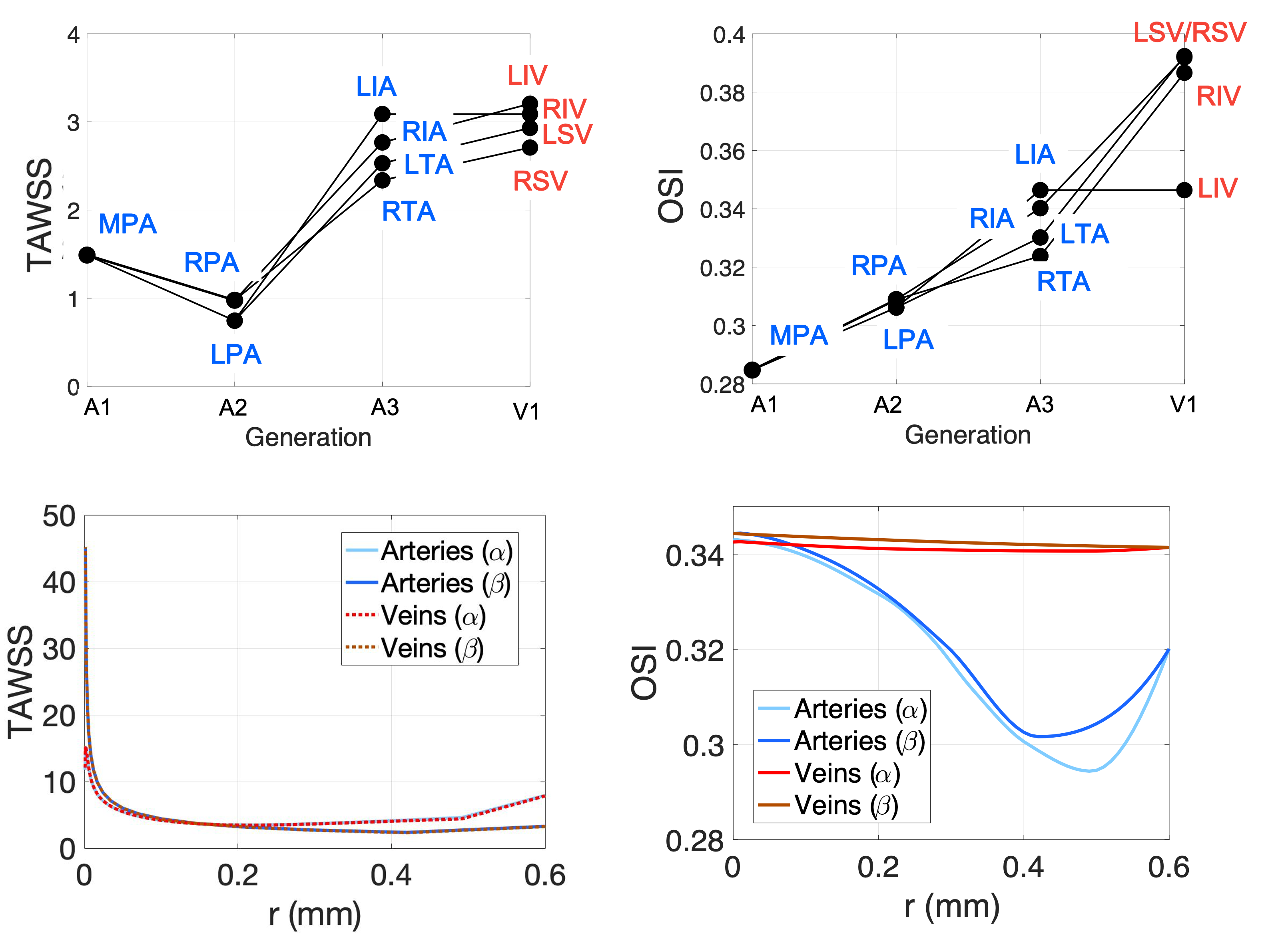}
    \caption{Time averaged wall shear stress (TAWSS) and oscillatory shear index (OSI) predictions for large and small arteries. Results for the large vessels is shown in the top row. Here, nodes represent the midpoint of the vessel, length of the edges represent distance between the midpoints of successive vessels, and terminal nodes represent veins. Results for TAWSS and OSI for the $\alpha$ and $\beta$ branches of the small arteries and veins are shown in the bottom row.}
    \label{fig:TAWSS_OSI}
\end{figure}

\section{Discussion}
\label{sec:discussion}
\par This study is the first to develop a multiscale model predicting WSS and CS in both large and small arteries and veins in the healthy adult human pulmonary circulation. Our results show that the large arteries have higher pressure and flow than the large veins, as well as more pronounced changes in area throughout the cardiac cycle. In the small vessels, we see that as vessel radius decreases, pressure decreases for arterioles but increases for venules. As vessel radius decreases, we also found that the volumetric flow decreases. Simulations show that WSS, TAWSS, and OSI generally increase in both large and small vessels as radius decreases. CS decreases in arterioles as radius decreases while increasing in venules as radius decreases. 

\par Our predictions of pressure, flow, and area in the large arteries and veins generally agree with the literature \cite{qureshi2014numerical}. There is a greater flow through the RPA than the LPA because the RPA has a larger radius than the LPA; however, we do not simulate curvature in vessels that affect in vivo distribution. We also note that the flow through the RSV is larger than the flow through the other veins because it is connected to the RTA, which has a larger radius (0.80 cm) than the other terminal arteries.  The structured tree model requires the diameters of pairs of large arteries and veins to be the same, thus the flow should be the largest through the RSV. The smallest venules have the largest pressure, which is still much less than the pressure in the arterioles. Veins have much lower pressure than arteries, which is depicted in our results and in literature \cite{gao2005role}. In Gao and Raj's study, they determined that under different levels of perfusate hematocrit and apparent viscosity, the large pulmonary arteries have higher pressure than the large pulmonary veins \cite{gao2005role}. The range in the value of pressure throughout the cardiac cycle agrees with Hall, who reported that the mean pulmonary arterial pressure should be 16 mmHg \cite{hall2020guyton}.  Chaliki et. al measured pulmonary arterial and venous pressure in mongrel dogs and found that pulmonary venous pressure was lower than pulmonary aterial pressure but higher than left atrial pressure. This is expected due to the veins location between the pulmonary artery and left atrium \cite{chaliki2002pulmonary}.  In our model, left atrial pressure is fixed at 2 mmHg; however, our results agree with that of Chaliki et. al.

\par Several studies have used CFD models to predict WSS, TAWSS, and OSI in large pulmonary arteries \cite{kheyfets2015patient,tang2012wall,yang2019evolution,zambrano2018image}. However, there has yet to be a study that investigates how local WSS changes in the entire pulmonary vasculature \cite{roux2020fluid}. The predictions we make of  these mechanical quantities for the large pulmonary arteries agree with what has been reported in these prior studies. In Zambrano et. al, it was reported that TAWSS values are higher in the third generation of the pulmonary arteries as opposed to the first and second generations \cite{zambrano2018image}. We see similar results, as our TWASS for the LIA, LTA, RIA, and RTA ranges between 2 and 3, whereas this quantites is between 0.75 and 1.5 for the MPA, RPA, and LPA, consistent with the results in Zambrano et. al  \cite{zambrano2018image}. In Yang et. al, TAWSS and OSI are computed in the arterial tree for control and PH patients. They find that in control subjects, the MPA has the lowest TAWSS and the TAWSS increases as vessel size decreases \cite{yang2019evolution}. Their results indicate that patients with severe PH tend to have lowered TAWSS throughout the arterial tree, as a result of decreased cardiac output and to less disturbed flow in smaller vessel branches \cite{yang2019evolution}. Results from Kheyfits et. al also indicate that WSS increases drastically in vessels with smaller diameter. Although they bound their WSS calculations to $50$ dyn/cm$^2$, they demonstrate that WSS is lowest in the MPA and increases as flow propagates down the arterial tree \cite{kheyfets2015patient}.  
Our study is the first to predict local WSS in a multiscale manner, by investigating this quantity in both large and small arteries and veins. This is a significant development because PH-LHF begins in the left ventricle and propagates to the right ventricle through the pulmonary veins, venules, capillaries, arterioles, and large pulmonary arteries. We study these phenomena in the healthy vasculature; however, by computing shear stress in both the microvasculature and the large pulmonary vessels, we are able to advance our understanding of how mechanical forces propagate throughout the pulmonary circulation. This in turn will give us a basis to investigate how PH-LHF develops and increases in severity in these regions. 

\par Most prior studies calculate WSS, TAWSS, and OSI using 3D simulations, but here we use a 1D model to make predictions. Nevertheless, our calculations yielded similar results with TAWSS ranging between 0 and 4 and OSI being near 0.3. In Zambrano et. al \cite{zambrano2018image}, TAWSS and OSI were reported with the normalized distance in MPA, RPA and LPA and we report the value at the midpoint of the vessels; however, our results for TAWSS and OSI are similar. In the large pulmonary arteries, they found that the MPA had elevated TAWSS and reduced OSI when compared to the smaller branches \cite{zambrano2018image}. We showed an analogous result with the MPA having larger TAWSS and smaller OSI than the RPA and LPA. Our predictions also examine these quantities in regions past the three large pulmonary arteries and we observe that the LIA, LTA, RIA, and RTA have both higher TAWSS and OSI than the MPA, RPA, and LPA. We also observe that RIV, RSV, LIV, and LSV have increased TAWSS and OSI. In contrast with our study, Yang et. al shows that in smaller branches of the pulmonary arteries, OSI is less than that for the MPA, RPA, and LPA \cite{yang2019evolution}. Discrepancies in OSI values could be due to data acquisition, geometry, the fact that Yang et. al studied 3D flow, and that they did not study arterioles. In general, Yang demonstrated that patients with severe PH had decreased OSI indicating that their blood flow is less oscillatory than healthy individuals \cite{yang2019evolution}. By characterizing OSI in the healthy pulmonary vasculature in both small and large vessels, we have a baseline value to compare to potential PH patients.

\par Other studies found that WSS and associated quantities are highest in the terminal arteries of the pulmonary vasculature, in agreement with our results. Yang et. al used morphometric trees to calculate WSS in distal pulmonary arteries to show that this quantity increases significantly with decreasing vessel size \cite{yang2019evolution}. They demonstrated that the mean WSS for vessel segments between $100-500 \mu m$ increased to $116$ dyn/cm$^2$ and that increased disease severity led to decreases in WSS values \cite{yang2019evolution}. Postles et. al \cite{postles2014dynamic} studied WSS in different disease conditions corresponding to PH and found that it was the highest in terminal arteries and increased with disease. As the disease progressed, they observed that the values of WSS climbed the arterial tree towards the proximal vessels, suggesting a link between WSS and vascular remodeling in PH, which is known to begin in small arteries and spread to larger arteries as the disease progresses \cite{postles2014dynamic}.  We also find that the OSI for venules is higher than the OSI for arterioles. 

\par To our knowledge, no previous studies have calculated CS in the arterioles and venules using a CFD model. Our results demonstrate that arterioles have a larger change in CS as radius decreases than venules. Because the veins are stiffer than arteries, the area does not deform as much over a cardiac cycle. This leads to a more dramatic change in CS for arteries than veins. In future work, we will further investigate CS to investigate whether PH-LHF conditions lead to altered CS. 

\subsection{Future Developments and Limitations}
In the present study, we utilized a 1D CFD model that connects large arteries and veins with geometry from segmented images to small vessels represented by structured trees. We only study conditions that correspond to a healthy adult with a inflow profile that has a normotensive cardiac output and left atrial pressure of 2 mmHg. We plan to further develop this study by including conditions that correspond to PH-LHF (elevated left atrial pressure, decreased flow, and increased vessel stiffness.) Analyzing results that correspond to disease progression will allow us to further understand the mechanisms involved in PH progression.

\par Another limitation of this study is that we neglect the pulmonary capillaries. Currently, flow in the smallest arterioles and venules (vessels with radius less than 0.001 mm) is not modeled. Future studies will combine three models: large arteries and veins with geometry from segmented images, small vessels represented by structured trees, and capillaries represented by a sheet model \cite{fung1969theory}.

\par Another limitation in this study is that we do not account for the gravitational gradient that impacts hemodynamics in the veins. Differences in hydrostatic pressures impact distensibility of vessels, vascular resistance, and blood flow  \cite{wieslander2019supine}. Because pulmonary veins lack smooth muscle, the vessel wall compliance can be expected to be high and hence, blood volume in the veins are sensitive to pressure changes caused by the effects of gravity \cite{wieslander2019supine}. 

\par Finally, we use a 1D model to generate our predictions, but WSS and CS are three-dimensional quantities. Since secondary flows developed in curved and bifurcating vessels cannot be simulated with a 1D model, WSS and WSS gradients cannot be computed as precisely \cite{grinberg2011modeling}. However, differences between subjects are likely greater than the precision of these 3D predictions.

\section{Conclusion}
The in-silico computational fluid dynamics model presented in this study is an important tool that provides a new way to analyze and investigate hypotheses related to understanding the physiological mechanisms underlying the progression of pulmonary vascular diseases. We are the first study to predict WSS and CS in both the large and small pulmonary arteries and veins, quantities that are hypothesized to alter mechanical forces that lead to PH-LHF. Prior to this study, these forces had not yet been quantitatively defined in the complete pulmonary vasculature for a healthy subject.  Our results for a physiologically healthy subject agree with previous studies, while also providing quantitative measurements for values not previously studied. 

\label{sec:conclusion}

\begin{acknowledgements}
This work was supported in part by the National Institute of Health (NIH-HL147590-01, NIH-NIAID 1R01AI139085-01), the National Science Foundation (NSF-DMS 1615820) and the American Heart Association (AHA 19PRE34380459).
\end{acknowledgements}

\bibliography{references.bib}{}

\begin{thebibliography}{10}
\providecommand{\url}[1]{{#1}}
\providecommand{\urlprefix}{URL }
\expandafter\ifx\csname urlstyle\endcsname\relax
  \providecommand{\doi}[1]{DOI~\discretionary{}{}{}#1}\else
  \providecommand{\doi}{DOI~\discretionary{}{}{}\begingroup
  \urlstyle{rm}\Url}\fi

\bibitem{antiga2008image}
Antiga, L., Piccinelli, M., Botti, L., Ene-Iordache, B., Remuzzi, A., Steinman,
  D.A.: An image-based modeling framework for patient-specific computational
  hemodynamics.
\newblock Medical \& biological engineering \& computing \textbf{46}(11), 1097
  (2008)

\bibitem{barron2007effect}
Barron, V., Brougham, C., Coghlan, K., McLucas, E., O’Mahoney, D.,
  Stenson-Cox, C., McHugh, P.E.: The effect of physiological cyclic stretch on
  the cell morphology, cell orientation and protein expression of endothelial
  cells.
\newblock Journal of materials science: Materials in medicine \textbf{18}(10),
  1973--1981 (2007)

\bibitem{birukov2009cyclic}
Birukov, K.G.: Cyclic stretch, reactive oxygen species, and vascular
  remodeling.
\newblock Antioxidants \& redox signaling \textbf{11}(7), 1651--1667 (2009)

\bibitem{bleakley2015endothelial}
Bleakley, C., Hamilton, P.K., Pumb, R., Harbinson, M., McVeigh, G.E.:
  Endothelial function in hypertension: victim or culprit?
\newblock The Journal of Clinical Hypertension \textbf{17}(8), 651--654 (2015)

\bibitem{chaliki2002pulmonary}
Chaliki, H.P., Hurrell, D.G., Nishimura, R.A., Reinke, R.A., Appleton, C.P.:
  Pulmonary venous pressure: relationship to pulmonary artery, pulmonary wedge,
  and left atrial pressure in normal, lightly sedated dogs.
\newblock Catheterization and cardiovascular interventions \textbf{56}(3),
  432--438 (2002)

\bibitem{chambers2020morphometry}
Chambers, M.J., Colebank, M.J., Qureshi, M.U., Clipp, R., Olufsen, M.S.:
  Structural and hemodynamic properties of murine arterial networks under
  hypoxia-induced pulmonary hypertension.
\newblock Proc Inst Mech Eng Part H: J Eng Med p. 0954411920944110 (2020).
\newblock \doi{10.1177/0954411920944110}

\bibitem{davies2009hemodynamic}
Davies, P.F.: Hemodynamic shear stress and the endothelium in cardiovascular
  pathophysiology.
\newblock Nature clinical practice Cardiovascular medicine \textbf{6}(1),
  16--26 (2009)

\bibitem{van2019transition}
van Duin, R.W., Stam, K., Cai, Z., Uitterdijk, A., Garcia-Alvarez, A., Ibanez,
  B., Danser, A.J., Reiss, I.K., Duncker, D.J., Merkus, D.: Transition from
  post-capillary pulmonary hypertension to combined pre-and post-capillary
  pulmonary hypertension in swine: a key role for endothelin.
\newblock The Journal of physiology \textbf{597}(4), 1157--1173 (2019)

\bibitem{fung1969theory}
Fung, Y., Sobin, S.: Theory of sheet flow in lung alveoli.
\newblock Journal of Applied Physiology \textbf{26}(4), 472--488 (1969)

\bibitem{gao2005role}
Gao, Y., Raj, J.U.: Role of veins in regulation of pulmonary circulation.
\newblock American Journal of Physiology-Lung Cellular and Molecular Physiology
  \textbf{288}(2), L213--L226 (2005)

\bibitem{gerges2015PHprogeression}
Gerges, M., Gerges, C., Pistritto, A., Lang, M., Trip, P., Jakowitsch, J.,
  Binder, T., Lang, I.: Pulmonary hypertension in heart failure. epidemiology,
  right ventricular function, and survival.
\newblock Am J Respir Crit Care Med \textbf{192}, 1234–1246 (2015).
\newblock \doi{10.1164/rccm.201503-0529OC}

\bibitem{ghio2001PH}
Ghio, S., Gavazzi, A., Campana, C., Inserra, C., Klersy, C., Sebastiani, R.,
  Arbustini, E., Recusani F~amd~Tavazzi, L.: Independent and additive
  prognostic value of right ventricular systolic function and pulmonary artery
  pressure in patients with chronic heart failure.
\newblock J Am Coll Cardiol \textbf{37}, 183–188 (2001).
\newblock \doi{10.1016/S0735-1097(00)01102-5}

\bibitem{grinberg2011modeling}
Grinberg, L., Cheever, E., Anor, T., Madsen, J.R., Karniadakis, G.: Modeling
  blood flow circulation in intracranial arterial networks: a comparative 3d/1d
  simulation study.
\newblock Annals of biomedical engineering \textbf{39}(1), 297--309 (2011)

\bibitem{guazzi2012PHmorbidity}
Guazzi, M., Borlaug, B.: Pulmonary hypertension due to left heart disease.
\newblock Circulation \textbf{126}, 975–990 (2012).
\newblock \doi{0.1161/CIRCULATIONAHA.111.085761}

\bibitem{guglin2010PHmorbidity}
Guglin, M., Khan, H.: Pulmonary hypertension in heart failure.
\newblock J Card Fail \textbf{16}, 461–474 (2010).
\newblock \doi{10.1016/j.cardfail.2010.01.003}

\bibitem{hall2020guyton}
Hall, J.E., Hall, M.E.: Guyton and Hall textbook of medical physiology e-Book.
\newblock Elsevier Health Sciences (2020)

\bibitem{huang1996morphometry}
Huang, W., Yen, R., McLaurine, M., Bledsoe, G.: Morphometry of the human
  pulmonary vasculature.
\newblock Journal of applied physiology \textbf{81}(5), 2123--2133 (1996)

\bibitem{kheyfets2015patient}
Kheyfets, V.O., Rios, L., Smith, T., Schroeder, T., Mueller, J., Murali, S.,
  Lasorda, D., Zikos, A., Spotti, J., Reilly~Jr, J.J., et~al.: Patient-specific
  computational modeling of blood flow in the pulmonary arterial circulation.
\newblock Computer methods and programs in biomedicine \textbf{120}(2), 88--101
  (2015)

\bibitem{lam2009PH}
Lam, C., Roger, V., Rodeheffer, R., Borlaug, B., Enders, F., Redfield, M.:
  Pulmonary hypertension in heart failure with preserved ejection fraction: a
  community-based stud.
\newblock J Am Coll Cardiol \textbf{53}, 1119–1126 (2009).
\newblock \doi{10.1016/j.jacc.2008.11.051}

\bibitem{Thomas2010}
Michel, T., Paul, M.V.: Cellular signaling and no production.
\newblock Pflugers Archiv : European journal of physiology \textbf{459}(6),
  807--816 (2010).
\newblock \doi{10.1007/s00424-009-0765-9}

\bibitem{miller2013PHprogression}
Miller, W., Grill, D., Borlaug, B.: Clinical features, hemodynamics, and
  outcomes of pulmonary hypertension due to chronic heart failure with reduced
  ejection fraction: pulmonary hypertension and heart failure.
\newblock JACC Heart Fail \textbf{1}, 290–299 (2013).
\newblock \doi{10.1016/j.jchf.2013.05.001}

\bibitem{moraes1997pulmonary}
Moraes, D., Loscalzo, J.: Pulmonary hypertension: newer concepts in diagnosis
  and management.
\newblock Clinical cardiology \textbf{20}(8), 676--682 (1997)

\bibitem{mozaffarian2016LHF}
Mozaffarian, D., Benjamin, E., Go, A., Arnett, D., Blaha, M., Cushman, M., Das,
  S., de~Ferranti, S., Despres, J., Fullerton, H., Howard, V., Huffman, M.,
  Isasi, C., Jimenez, M., Judd, S., Kissela, B., Lichtman, J., Lisabeth, L.,
  Liu, S., Mackey, R., Magid, D., McGuire, D., Mohler, E.r., Moy, C., Muntner,
  P., Mussolino, M., Nasir, K., Neumar, R., Nichol, G., Palaniappan, L.,
  Pandey, D., Reeves, M., Rodriguez, C., Rosamond, W., Sorlie, P., Stein, J.,
  Towfighi, A., Turan, T., Virani, S., Woo, D., Yeh, R., Turner, M., Members,
  W.G., Committee, A.H.A.S., Subcommittee, S.S.: Heart disease and stroke
  statistics-2016 update: a report from the american heart association.
\newblock Circulation \textbf{133}, e38–e360 (2016).
\newblock \doi{10.1161/CIR.0000000000000350}

\bibitem{olufsen2000numerical}
Olufsen, M.S., Peskin, C.S., Kim, W.Y., Pedersen, E.M., Nadim, A., Larsen, J.:
  Numerical simulation and experimental validation of blood flow in arteries
  with structured-tree outflow conditions.
\newblock Annals of biomedical engineering \textbf{28}(11), 1281--1299 (2000)

\bibitem{papadaki1999quantitative}
Papadaki, M., Mclntire, L.V.: Quantitative measurement of shear-stress effects
  on endothelial cells.
\newblock In: Tissue Engineering Methods and Protocols, pp. 577--593. Springer
  (1999)

\bibitem{paszkowiak2003arterial}
Paszkowiak, J.J., Dardik, A.: Arterial wall shear stress: observations from the
  bench to the bedside.
\newblock Vascular and endovascular surgery \textbf{37}(1), 47--57 (2003)

\bibitem{postles2014dynamic}
Postles, A., Clark, A.R., Tawhai, M.H.: Dynamic blood flow and wall shear
  stress in pulmonary hypertensive disease.
\newblock In: 2014 36th Annual International Conference of the IEEE Engineering
  in Medicine and Biology Society, pp. 5671--5674. IEEE (2014)

\bibitem{pries1992blood}
Pries, A.R., Neuhaus, D., Gaehtgens, P.: Blood viscosity in tube flow:
  dependence on diameter and hematocrit.
\newblock American Journal of Physiology-Heart and Circulatory Physiology
  \textbf{263}(6), H1770--H1778 (1992)

\bibitem{qureshi2014numerical}
Qureshi, M.U., Vaughan, G.D., Sainsbury, C., Johnson, M., Peskin, C.S.,
  Olufsen, M.S., Hill, N.: Numerical simulation of blood flow and pressure drop
  in the pulmonary arterial and venous circulation.
\newblock Biomechanics and modeling in mechanobiology \textbf{13}(5),
  1137--1154 (2014)

\bibitem{ramu2016PHmorbidity}
Ramu, B., Thenappan, T.: Evolving concepts of pulmonary hypertension secondary
  to left heart disease.
\newblock Curr Heart Fail Rep \textbf{13}, 92–102 (2016).
\newblock \doi{10.1007/s11897-016-0284-x}

\bibitem{ravi2013PHprogeression}
Ravi, Y., Selvendiran, K., Naidu, S.K., Meduru, S., Citro, L.A., Bognar, B.,
  Khan, M., Kalai, T., Hideg, K., Kuppusamy, P., Sai-Sudhakar, C.B.: Pulmonary
  hypertension secondary to left-heart failure involves peroxynitrite-induced
  downregulation of pten in the lung.
\newblock Hypertension \textbf{61}(3), 593–601 (2013).
\newblock \doi{10.1161/HYPERTENSIONAHA.111.00514}

\bibitem{reymond2011validation}
Reymond, P., Bohraus, Y., Perren, F., Lazeyras, F., Stergiopulos, N.:
  Validation of a patient-specific one-dimensional model of the systemic
  arterial tree.
\newblock American Journal of Physiology-Heart and Circulatory Physiology
  \textbf{301}(3), H1173--H1182 (2011)

\bibitem{reymond2012patient}
Reymond, P., Perren, F., Lazeyras, F., Stergiopulos, N.: Patient-specific mean
  pressure drop in the systemic arterial tree, a comparison between 1-d and 3-d
  models.
\newblock Journal of biomechanics \textbf{45}(15), 2499--2505 (2012)

\bibitem{riches1973blood}
Riches, A., Sharp, J., Thomas, D.B., Smith, S.V.: Blood volume determination in
  the mouse.
\newblock The Journal of physiology \textbf{228}(2), 279--284 (1973)

\bibitem{roux2020fluid}
Roux, E., Bougaran, P., Dufourcq, P., Couffinhal, T.: Fluid shear stress
  sensing by the endothelial layer.
\newblock Frontiers in Physiology \textbf{11} (2020)

\bibitem{selzer1992understanding}
Selzer, A.: Understanding heart disease.
\newblock Univ of California Press (1992)

\bibitem{simonneau2019PHdiagnosis}
Simonneau, G., Montani, D., Celermajer, D., Denton, C., Gatzoulis, M., Krowka,
  M., Williams, P., Souza, R.: Haemodynamic definitions and updated clinical
  classification of pulmonary hypertension.
\newblock Eur Respir J \textbf{53}, 1801,913 (2019).
\newblock \doi{10.1183/13993003.01913-2018}

\bibitem{tang2012wall}
Tang, B.T., Pickard, S.S., Chan, F.P., Tsao, P.S., Taylor, C.A., Feinstein,
  J.A.: Wall shear stress is decreased in the pulmonary arteries of patients
  with pulmonary arterial hypertension: an image-based, computational fluid
  dynamics study.
\newblock Pulmonary circulation \textbf{2}(4), 470--476 (2012)

\bibitem{wieslander2019supine}
Wieslander, B., Ramos, J.G., Ax, M., Petersson, J., Ugander, M.: Supine, prone,
  right and left gravitational effects on human pulmonary circulation.
\newblock Journal of Cardiovascular Magnetic Resonance \textbf{21}(1), 1--15
  (2019)

\bibitem{yang2019evolution}
Yang, W., Dong, M., Rabinovitch, M., Chan, F.P., Marsden, A.L., Feinstein,
  J.A.: Evolution of hemodynamic forces in the pulmonary tree with
  progressively worsening pulmonary arterial hypertension in pediatric
  patients.
\newblock Biomechanics and modeling in mechanobiology \textbf{18}(3), 779--796
  (2019)

\bibitem{zambrano2018image}
Zambrano, B., McLean, N., Zhao, X., Tan, J.L., Zhong, L., Figueroa, C., Lee,
  L., Baek, S.: Image-based computational assessment of vascular wall mechanics
  and hemodynamics in pulmonary arterial hypertension patients.
\newblock J Biomech \textbf{68}, 84--92 (2018)

\bibitem{zamir2002physics}
Zamir, M., Budwig, R.: Physics of pulsatile flow.
\newblock Appl. Mech. Rev. \textbf{55}(2), B35--B35 (2002)

\end{thebibliography}
\bibliographystyle{spmpsci}  

\end{document}